\documentstyle[prb,epsfig,preprint,aps,amsmath]{revtex}

\begin{document}

\draft

\title{Testing of two-dimensional local approximations in the current-spin and
spin-density-functional theories.}

\author{H.~Saarikoski, E. R\"as\"anen, S. Siljam\"aki, A.~Harju,
M.~J.~Puska, and R.~M.~Nieminen}

\address{Laboratory of Physics, Helsinki University of Technology,
P.O. Box 1100, FIN-02015 HUT, FINLAND} 

\date{\today}
\maketitle

\begin{abstract}
We study a model quantum dot system in an external magnetic field by using 
both the spin-density-functional theory and the current-spin-density-functional
theory. The theories are used with local approximations
for the spin-density and the vorticity.
The reliabilities of different parametrizations for the 
exchange-correlation functionals are tested by comparing the ensuing
energetics with quantum Monte Carlo results. The limit where the vorticity 
dependence should be used in the exchange-correlation functionals is 
discussed.
\end{abstract}


\section{Introduction}
\label{sec:introduction}


The success of the spin-density-functional theory (SDFT) \cite{dreizlergross}
within  the local-spin-density approximation (LSDA) to predict
accurate results for
the electronic structure of two-dimensional quantum dot systems depends
on the exchange-correlation functionals used.
Until recently, functionals based on the diffusion quantum Monte Carlo (DMC)
calculations by Tanatar and Ceperley \cite{tanatar} have been widely employed.
The simulations were performed for the spin-compensated and spin-polarized
two-dimensional  electron gas (2DEG).
Recently, Attaccalite {\em et al.} \cite{attaccalite,gorigiorgi}
made fixed-node DMC calculations with improved accuracy for the 2DEG including 
several partial spin-polarizations. They provided interpolation
formulas which fulfil several exact results at the low and high
electron density limits.

In the current-spin-density-functional theory (CSDFT), used to describe
electron systems in magnetic fields, the exchange-correlation energy depends
also on the paramagnetic current density \cite{vignalerasolt}. In the 
local approximation the current-dependence is converted to the 
dependence on the vorticity or on the Landau-level filling factor. 
The low filling factor (strong magnetic field) limit for the
totally spin-polarized electron gas, is well known 
from the works by Levesque, Weis and MacDonald \cite{levesque}, Fano and 
Ortolani \cite{fano}, and Price and Das Sarma \cite{rodney}. The problem
is how to interpolate the exchange-correlation energy between this limit 
and the high filling factor (zero-magnetic field) limit for a given 
spin-polarization. Several interpolation schemes have been suggested 
\cite{rasolt,heinonen,koskinen}.

The purpose of the present work is to study the reliabilities of the two
zero-field exchange-correlation functionals and
various exchange-correlation interpolation schemes with respect to the
filling factor in the CSDFT calculations. We use
a parabolically-confined quantum dot as the test system and
compare our SDFT and CSDFT total energies with those obtained in variational
quantum Monte Carlo (VMC) calculations \cite{VMC,saarikoski}.
In the zero-field case the parabolic confinement 
is lowered towards the limit of Wigner-crystallization
and the difference between the 
total energies of the spin-polarized and spin-compensated solutions are
monitored. For a finite confinement we calculate the ground state energy
as a function of the external magnetic field. Besides the CSDFT schemes,
calculations are performed also with the SDFT, {\em i.e} ignoring the 
current-dependence of the exchange-correlation energy. Thus, our calculations
enlighten how important improvement the CSDFT is in comparison
with the SDFT. For the quantitative comparisons we need numerically
accurate results. We perform the SDFT and CSDFT calculations on two-dimensional
point grids imposing no symmetry restrictions \cite{saarikoski,heiskanen}.
The convergence with respect to the grid size and other numerical
approximations is carefully tested.
The VMC results are converged beyond the statistical noise.

The outline of the present paper is as follows. In Sec. \ref{zero-field}
we shortly describe the exchange-correlation functionals
in zero magnetic field 
and give our results for a six-electron quantum dot as a function of the
confinement. In Sec. \ref{finite-field} we discuss the interpolation as
a function of the Landau level filling factor and show our results
for the magnetic field dependence on the ground-state total energy. 
Sec. \ref{conclusions} contains the conclusions.
Effective atomic units are used in the formulas throughout the paper and
for presenting the results 
they are converted by using the material parameters of GaAs, {\em i.e.}
the dielectric constant $\epsilon=12.4$, the effective mass $m^*= 0.067$ and
the gyromagnetic constant $g^*= -0.44$.
We choose the coordinate system so that $x$ and $y$ are in the
plane of the dot and the $z$-axis is perpendicular to the plane.

\section{zero-field and the low-density limit}
\label{zero-field}

Attaccalite {\em et al.} (AMGB) \cite{attaccalite,gorigiorgi} calculated 
the ground state of the 2DEG with the fixed-node DMC.
Compared to the work by Tanatar and Ceperley (TC)\cite{tanatar}
there are a number of improvements in the numerical calculations.
Backflow correlations in many-body wave functions are included,
and infinite size extrapolations are performed in the Monte Carlo data. 
An important new feature of the results by
Attaccalite {\em et al.} is the appearence of a spin-polarized ground state 
before the Wigner crystallization. 

On the basis of their DMC calculations, AMGB \cite{attaccalite}
proposed a new analytic representation of the correlation energy.
It takes into account several exact results 
in the low and high density limits. 
They give the exchange-correlation energy of the homogeneous electron gas as
\begin{equation}
e_{\rm xc}(r_s,\zeta)=e_x(r_s,\zeta)+(e^{-\beta r_s}-1)e_{\rm x}^{(6)}(r_s,\zeta)
+\alpha_0(r_s)+\alpha_1(r_s)\zeta^2+\alpha_2(r_s)\zeta^4.\label{zero}
\end{equation}
Here $r_s=1/\sqrt{\pi n}$ is the density parameter, calculated from the 
two-dimensional electron density $n$, and $\zeta=(n_\uparrow - n_\downarrow)/n$
is the spin polarization determined by the spin-up and spin-down electron densities
$n_\uparrow$ and $n_\downarrow$, respectively. $e_x(r_s,\zeta)$ is the 
exchange energy, which can be written as
\begin{equation}
e_{\rm x}(r_s,\zeta)=-2\sqrt{2}[(1+\zeta)^{3/2}+(1-\zeta)^{3/2}]/3\pi r_s.\label{x}
\end{equation}
In Eq. (\ref{zero}), $e^{(6)}_{\rm x}$ contains the terms of the Taylor expansion of 
$e_{\rm x}$ with respect to $\zeta$ at $\zeta=0$ which are
beyond the fourth order in $\zeta$,
$\alpha$'s are density dependent functions of the generalized Perdew-Wang
form\cite{perdew}, and $\beta=1.3386$.

The above form (\ref{zero}) is based on DMC calculations for partial
polarizations ($0 < \zeta < 1$). TC \cite{tanatar} made
DMC calculations only for spin-compensated ($\zeta = 0$) and spin-polarized
($\zeta = 1$) systems. To calculate the correlation potential for finite
polarizations the TC data is often used with the exchange-like
interpolation\cite{koskinen2} of Eq. (\ref{x}).
This leads to deviations from the
exact results at high electron densities and too small 
spin-susceptibilities at low densities \cite{attaccalite}. 

Our test system is a six-electron quantum dot confined by a
parabolic external potential
\begin{equation}
V_{\rm ext}={1 \over 2} m^* \omega^2_0 r^2,
\end{equation}
where $\hbar \omega_0$ is the confinement strength. 
First we use a zero magnetic field and compare the results
obtained with the two LSDA functionals to the VMC results.
We study the weak-confinement limit at which 
the electron density in the dot is low and therefore the contribution 
of the exchange-correlation energy to the total energy is relatively
large. Thus, the high-correlation effects in this particular 
system, {\em i.e.} the Wigner crystallization, spin density wave (SDW) 
formation, and spin-polarization, are assumed to be sensitive
to the LSDA functional used.

Fig. 1 shows the energy difference between the $S_z=3$ and 
$S_z=0$ spin states as a function of the confinement 
strength. The agreement with the VMC results is remarkably better for the AMGB 
functional than for the TC parametrization. In the AMGB results
the spin-polarization is favoured at confinement strengths below
$\hbar\omega_0=0.23$ meV. This is roughly in the middle between the TC 
and VMC results of 0.18 and 0.28 meV, respectively. As the confinement 
is made stronger, the difference between the parametrizations grows 
further enhancing the advantage of the new LSDA so that it smoothly 
follows the VMC curve. The origin of the difference is in the total 
energy of the polarized ($S_z=3$) state. It is lowered when the
AMGB functional is used instead of the TC one. This observation is in
accord with the result of Gori-Giorgi {\em et al.} \cite{gorigiorgi}
that the improvement due to the new functional is directly proportional 
to the polarization and the electron density of the system. In the present 
case, however, there is a  significant difference between these two 
approximations down to $\hbar\omega_0\simeq{0.15}$ meV. This corresponds to 
quite a small electron density, {\em i.e.}
$r_s=(N^{1/4}\hbar\omega_0)^{-2/3} \sim 14$ (Ref. \onlinecite{koskinen2}).
In our calculations,
this is actually the smallest confinement strength for which we have
strictly converged SDFT results.

The transition point to the $S_z=0$  SDW state also seems to depend 
strongly on the applied type of LSDA. By using the TC functional 
we find the breaking of the spin symmetry at $\hbar\omega_0\simeq{0.45}$ meV, 
whereas with the new LSDA the transition occurs already at 
$\hbar\omega_0\simeq{0.8}$ meV, which corresponds to $r_s=4.5$.
This should be compared with the estimation
by Koskinen {\em et al.} that in the case of closed shells,
a SDW is found for $r_s\gtrsim{5}$ (Ref. \onlinecite{koskinen2}). 
Fig. 1 shows also the results obtained by forcing the spin densities to be
equal, {\em i.e.} preventing the SDW formation. It can be seen 
that the values of the SDW solutions are in a clearly better agreement 
with the VMC results than those of the symmetry-restricted solution. 
The relative amplitude of the SDW grows rapidly at low confinements below 
the transition point, and the electron density shows localization around 
six maxima, corresponding to the Wigner crystallization. The behavior of 
the electron density is presented in more detail in
Ref. \onlinecite{saarikoski}.


\section{External magnetic field}
\label{finite-field}

Next we add an external magnetic field perpendicular to the dot ({\em i.e.}
parallel
to the $z$-axis) and fix the confinement strength to $5\;{\rm meV}$.
SDFT calculations are done 
with a minimal substitution of the external vector potential
in the Schr\"odinger equation,
and by using the zero-field exchange-correlation functionals for the 2DEG.
In the CSDFT \cite{vignalerasolt} the exchange-correlation functionals depend 
on the electron currents in the system, and they are functionals of the
spin densities and the vorticity,
\begin{equation}
\gamma({\bf r}) = \nabla \times {{\bf j}_p({\bf r}) \over n({\bf r})}\biggr\vert_z.
\label{vorti}
\end{equation}
Above, $n$ is the electron density and
${\bf j}_p$ is the paramagnetic current density.
Equally, we can use the Landau level filling factor given by
\begin{equation}
\nu = 2 \pi n/\gamma.
\end{equation}

The data for the 2DEG in magnetic field is scarce. Fano and Ortolani \cite{fano} 
used the Monte Carlo data by Morf {\em et al.} \cite{morf} and gave
the $\nu$ dependence 
of the 2DEG exchange-correlation energy for low $\nu$ values from 0 to 
about 0.8, corresponding to high magnetic fields. At the limit of the infinite 
magnetic field ($\nu \rightarrow 0$), the resulting curve agrees well with the
results by Levesque, Weis, and MacDonald \cite{levesque}.
Price and Das Sarma\cite{rodney} obtained for the polarized 2DEG
the density dependence of the exchange-correlation energy at the
$\nu$ values of 1/7, 1/5, 1/3, 1 and 2. In the low $\nu$ region 
the fractional quantum Hall effect causes cusps in the exchange-correlation 
energy. Heinonen {\em et al.} \cite{heinonen} took them into account
in their modeling, but we have ignored them because in our test case of
the six-electron quantum dot, the $\nu$ values were above 0.9
in the magnetic fields up to 10 T.

In the present calculations we have used 
expressions given by Koskinen {\em et al.}\cite{koskinen} 
and Ferconi and Vignale \cite{ferconi} for the magnetic-field dependence of
the exchange-correlation (See also Ref. \onlinecite{rasolt}).
Koskinen {\em et al.} fitted their functionals 
in the high vorticity limit and used the following formula for the
interpolation to the zero-field limit,
\begin{equation}
e^{\rm K}_{\rm xc}(n,\zeta,\nu) = -0.782\sqrt{2\pi n}e^{-f(\nu)}+e^{B=0}_{\rm xc}(n,\zeta)
(1-e^{-f(\nu)}),\label{koskinenexc}
\end{equation}
where $f(\nu)=1.5\nu+7\nu^4$. Koskinen {\em et al.} used the TC functional as 
the zero-field limit $e_{\rm xc}^{\rm B=0}$. We replace it
by the AMGB functional (Eq. (\ref{zero})).
Ferconi and Vignale applied a Pad\'e approximant, fitting the low $\nu$
limit of Levesque {\em et al.} \cite{levesque} to the zero magnetic field 
functionals, {\em i.e.},
\begin{equation}
e^{\rm PADE}_{\rm xc}(n,\zeta, \nu)= (e^{\rm LWM}_{\rm xc}(n,\nu) + \nu^4 e^{B=0}_{\rm xc}(n,\zeta))/(1+\nu^4),
\label{eq:pade}
\end{equation}
where $e^{\rm LWM}_{\rm xc}$ is the interpolation formula for the infinite
magnetic field limit \cite{levesque},
\begin{equation}
e^{\rm LWM}_{\rm xc}= -0.782133\sqrt{2\pi n}(1-0.211\nu^{0.74}+0.012\nu^{1.7}).
\label{LWMexc}
\end{equation}
As before, we insert the AMGB functional for $e_{\rm xc}^{\rm B=0}$.
The above formulae interpolate between the fully polarized electron
gas values at high magnetic fields and the zero-field limit, which may have
arbitrary polarization. Data for the intermediate polarizations at
high magnetic fields would be desirable to test the interpolation further.

Fig. \ref{fig:exc}(a) shows the quantum Monte Carlo data by Price and Das Sarma
\cite{rodney} and Fano and Ortolani\cite{fano} 
for the exchange-correlation energy of the polarized 2DEG at $n=0.138$ a.u. 
The horizontal line indicates the zero-field limit of AMGB.
Fig. \ref{fig:exc}(a)
shows also values from exchange-correlation energy by Koskinen {\em et al.}
(Eq. (\ref{koskinenexc})). At low $\nu$ values
all the data agree well but when $\nu$ increases the results by Price and
Das Sarma ($\nu=1$ and $\nu=2$) approach much more slowly
the zero-field limit than the data by Fano and Ortolani.
Therefore we propose a new functional which is an interpolation
to the data by Price and Das Sarma.
The resulting functional is 
\begin{equation}
e^{\rm{new}}_{\rm xc}(n,\zeta, \nu)= (e^{\rm LWM}_{\rm xc}(n,\nu) 
+ \nu^4 e^{B=0}_{\rm xc}(n,\zeta))/(1+0.0061\nu-0.0314\nu^2-0.0201\nu^3+\nu^4),
\label{eq:new}
\end{equation}
and it is shown in Fig. \ref{fig:exc}(b).
In the same figure we have also plotted $e^{\rm PADE}_{\rm xc}$
of Ferconi and Vignale. One should note that at $\nu < 1$ it approaches the 
zero-field limit much more slowly than the data by 
Fano and Ortolani in Fig. \ref{fig:exc}(a). The $e^{\rm PADE}_{\rm xc}$ shows
a maximum above the zero-field values, for which there is, however,
no physical reason.

The exchange-correlation potential $V_{\rm xc}$ is obtained in
the local approximation by calculating the functional derivative
\begin{equation}
V_{{\rm xc},\sigma}(n_\uparrow,n_\downarrow,\nu) =
\partial (ne_{\rm xc})/\partial n_\sigma.
\end{equation}
In the CSDFT, the vector potential depends on the derivative with respect to
the vorticity. The $x$ and $y$ components of the exchange-correlation
vector potential ${\bf A}_{\rm xc}$ are
\begin{equation}
A_{{\rm xc},x}=1/n {\partial \over \partial y} {\partial (ne_{\rm xc})\over \partial \gamma}, \label{axcx}
\end{equation}
\begin{equation}
A_{{\rm xc},y}=-1/n {\partial \over \partial x} {\partial (ne_{\rm xc})\over \partial \gamma}.\label{axcy}
\end{equation}
It should be noted that the contribution of the exchange-correlation energy 
to the total energy arises via the value of $e_{\rm xc}$
at the given density and vorticity {\em and} via its derivatives
with respect to both density and vorticity. Therefore the total energies
are sensitive to even slight variations in the exchange-correlation
functionals.

The vorticity is a functional of the ratio between the paramagnetic current 
density and the electron density (Eq. (\ref{vorti})). Its values increase 
rapidly towards low density areas causing numerical instabilities. To 
circumvent them we have used the convoluted form of Koskinen {\em et al.} 
\cite{koskinen} in the calculation of the partial derivatives 
$\partial e_{\rm xc}/\partial \gamma$ in Eqs. (\ref{axcx}) and (\ref{axcy}),
\begin{equation}
{\partial e_{\rm xc} \over \partial \gamma} \approx \int {(\gamma'-\gamma) \over
\sqrt{2 \pi} \Delta^3} e^{-(\gamma'-\gamma)^2 \over 2\Delta^2}
e_{\rm xc}(n,\zeta,\gamma').
\end{equation}
Above, the width $\Delta$ of the Gaussian function should be carefully
adjusted to the vorticity of the system at the given magnetic field strength.
Too small values for $\Delta$ may result in convergence problems.
On the other hand using too large a value for $\Delta$ may cause inaccuracy
in the results. In our calculations $\Delta$ ranges from 0.025 to 0.05.

Fig. \ref{fig:magfield} shows the total ground state energy
of a six-electron dot as a function of the magnetic field
up to 10 T. Results obtained with the SDFT and CSDFT formalisms
and different functionals are compared with the VMC data. 
Assuming again that the VMC results follow most faithfully the exact ones,
the CSDFT formalism is a clear improvement over the SDFT formalism:
The CSDFT results obtained with the AMGB functional using the Pad\'e approximant
for the filling factor interpolation, are closer to the VMC values than the
SDFT values already at magnetic fields slightly above 2 T. In this region
$S_z=0$.
The improvement of the CSDFT formalism
is also clear above  5 T, where $S_z=3$ and the maximum density droplet (MDD)
has been formed. The use of the AMGB functional instead of the TC functional
improves the SDFT results in the regions of large spin polarization, {\em i.e.},
around $B$ = 4.5T where $S_z=2$, and in the MDD region. This conclusion matches 
again well with the analysis by Gori-Giorgi {\em et al.} \cite{gorigiorgi}.

In the CSDFT, the ground state energy depends clearly on the chosen
exchange-correlation functional. As shown in Fig. 
\ref{fig:magfield} the interpolation 
suggested by Koskinen {\em et al.} (Eq. \ref{koskinenexc}) does not cause
a big difference between the CSDFT and SDFT results.
This is because $e^{\rm K}_{\rm xc}$ saturates quickly to the
zero-field values at about $\nu=1$ due to the exponential factors.
The clear difference obtained by using the Pad\'e approximant
(Eq. (\ref{LWMexc})) with an increasing magnetic field 
results in a much better agreement with the VMC data.
Second, the results obtained with the AMGB functional coincide well
in the MDD region with the VMC ones, whereas those obtained with the TC 
functional (not shown in Fig. \ref{fig:magfield}) are higher in energy
resembling the situation in the SDFT calculations.
Third, the new functional (Eq. (\ref{eq:new})) based on the data by
Price and Das Sarma seems to overcorrect the SDFT results.
We conclude that despite the good fit to the Monte Carlo data for polarized
2DEG, the derivatives of $e_{\rm xc}$ seem to be poorly approximated
by Eq.(\ref{eq:new}) as well as the spin compensated $\zeta=0$ values
for $e_{\rm xc}$.

In the MDD region of the VMC results the $S_z=2$ state constitutes 
the ground state in the CSDFT calculations  for the exchange-correlations
energies given in Eqs. (\ref{eq:pade}) and (\ref{eq:new}).
The corresponding energies 
are given in Fig. \ref{fig:magfield} as x-marks and dots, respectively.
However, we believe that the $S_z=2$ groundstate is an artifact due to 
the interpolation of the
exchange-correlation energy between the high-field totally-polarized and
the zero-field partially-polarized electron gas values.
Due to lack of simulation data the polarization dependence enters the
exchange-correlation functionals only through the
zero-field limits in Eqs. (\ref{koskinenexc}), (\ref{eq:pade}), and
(\ref{eq:new}). The high-field $S_z=2$ CSDFT ground states
have the same value for the total angular momentum,
$L_z=15$, as for the $S_z=3$ MDD-state, but the electron 
at the innermost orbital has a flipped spin.
Therefore, Hund's rule should favor the totally polarized $S_z=3$ state.
Further, we see that also around $B$ = 4.5T, where both the CSDFT and VMC
give the $S_z=2$ state, the CSDFT clearly underestimates the total energy.

\section{conclusions}
\label{conclusions}

We have studied the electronic structure of a parabolically-confined
six-electron quantum dot using the spin-density-functional and 
the current-spin-density-functional theories. In particular
the effects of weak confinement and strong external magnetic
field were considered. Our main aim was to investigate, how reliably the
different local approximations for the electron exchange and correlation
follow variational quantum Monte Carlo results, which we consider as
benchmarks. We have especially investigated the various interpolation schemes
between the spin-compensated and totally polarized electron gas and
the vorticity dependence of the exchange-correlation functionals.

We find that the new LSDA functional by Attaccalite {\em et al.} gives in 
the zero magnetic field much better results for the total energy than the 
old form by Tanatar and Ceperley. According to our results, the same is 
also true in an external magnetic field for a six-electron quantum dot with 
a typical confinement of $\hbar \omega_0$ = 5 meV. For this system the effect 
of currents on the exchange-correlation energy becomes important in magnetic 
fields larger than 2 - 3~T. In our calculations
for these high magnetic fields the local Landau level filling factor  $\nu$ is 
of the order of 0.9. The simulation data for the homogeneous electron
gas in this regime is scarce, which hampers establishing of 
interpolation forms between the low and high magnetic field limits
(large and small Landau level filling factors). The Pad\'e interpolation
by Ferconi and Vignale gives a very good
agreement with the quantum Monte Carlo results in the case of
spin-compensated and spin-polarized states of our system. The
exponential interpolation form by Koskinen {\em et al.} underestimates
clearly the effect of the magnetic field. Our results for the partially
spin-polarised states in magnetic field are remarkably worse than those 
for the spin-compensated and spin-polarized states. Thus, our tests
with the CSDFT call for further simulations for the homogeneous
electron gas in order to determine reliable interpolation forms for
the  vorticity and polarization dependence of the exchange-correlation 
functionals.

\acknowledgments

This work has been supported by Academy of Finland through the Centre
of Excellence Program (2000-2005).

\begin{figure}
\centerline{\psfig{figure=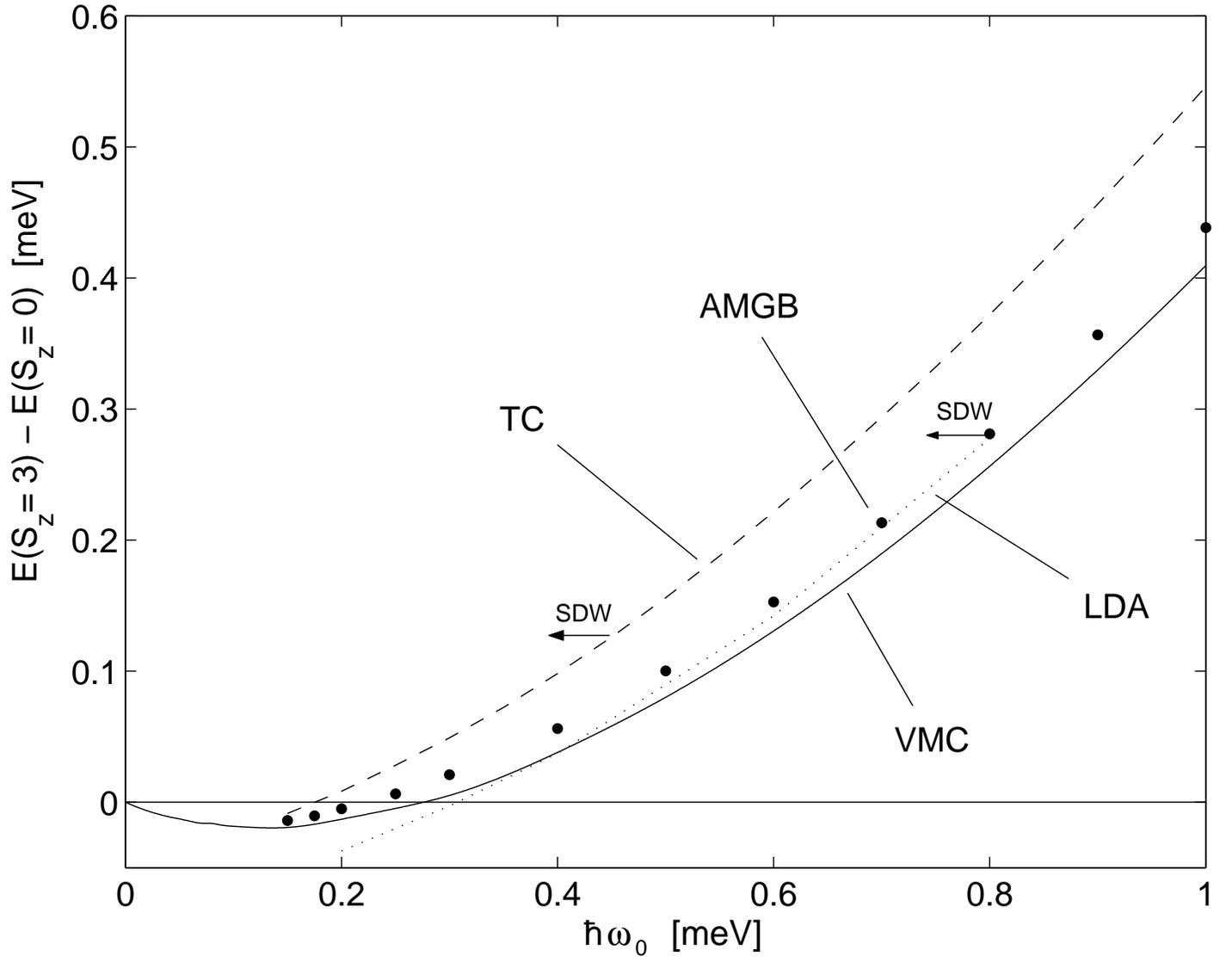,height=15cm}}
\vspace{1cm}
\caption{
Total energy difference between the $S_z=3$ and $S_z=0$ spin
states in a six-electron quantum dot. The results obtained with the
SDFT using the LSDA functionals by TC \cite{tanatar} and AMGB 
\cite{attaccalite} are compared with the VMC data. The onsets 
of SDW solutions with decreasing confinement are marked by arrows. 
The dotted curve shows the symmetry-restricted results, corresponding 
to the LDA.
}
\label{fig:weak}
\end{figure}

\begin{figure}
\centerline{\psfig{figure=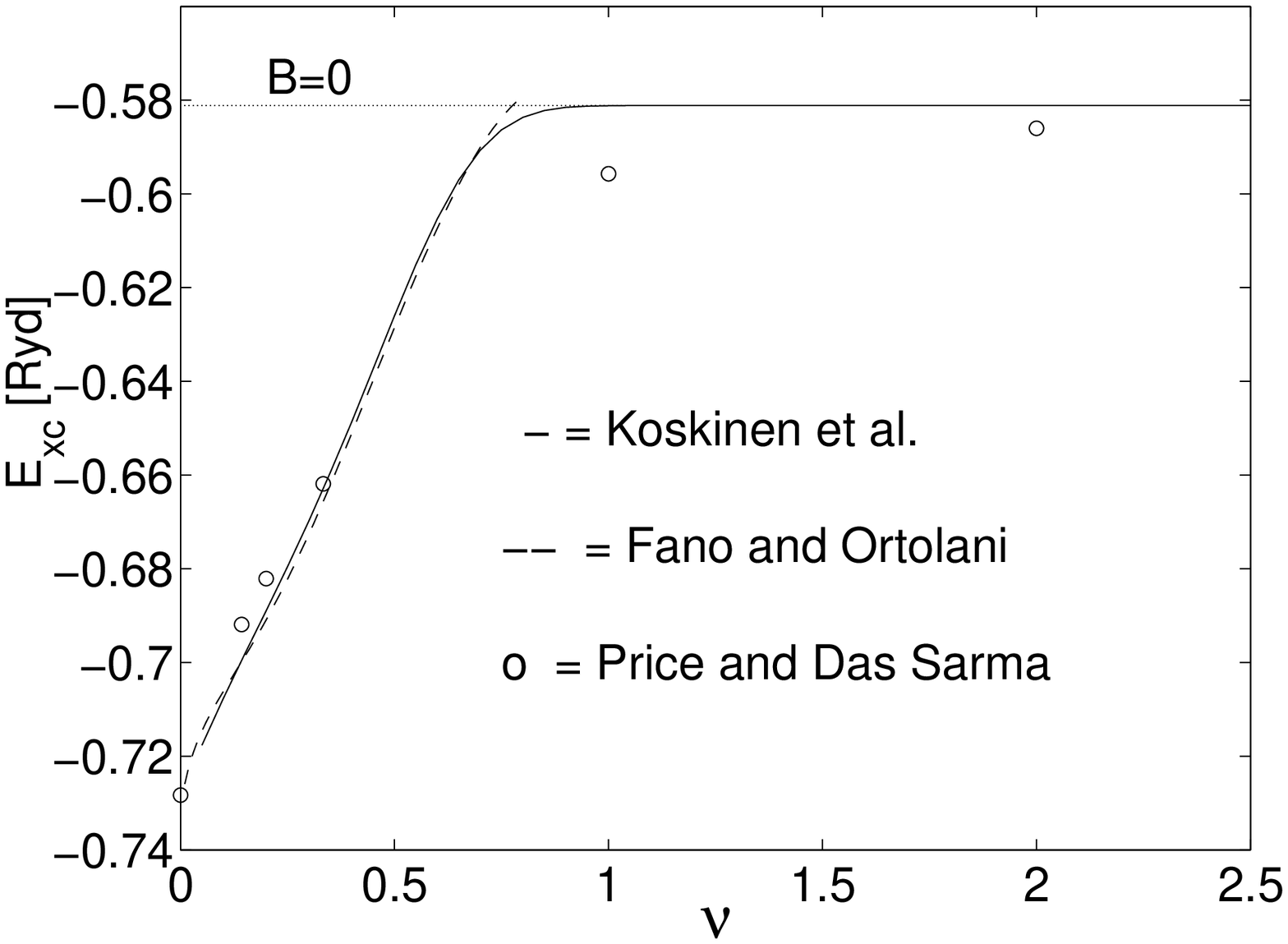,height=8cm}}
\centerline{\Huge (a)}
\vspace{0.5cm}
\centerline{\psfig{figure=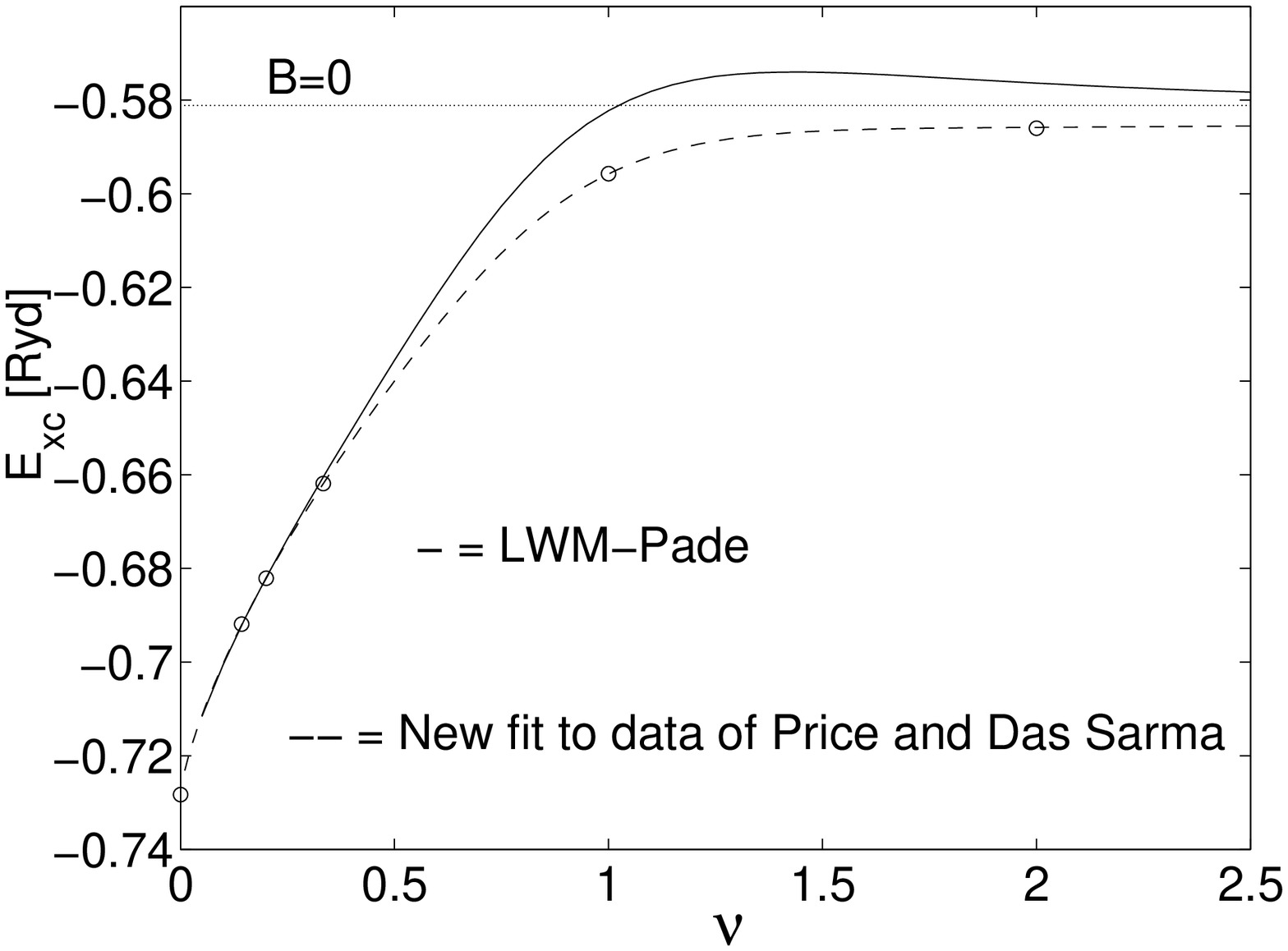,height=8cm}}
\centerline{\Huge (b)}
\vspace{0.5cm}
\caption{
Exchange-correlation energy per electron as a function of the
Landau level filling factor for the spin-polarized 2DEG with $n=0.138$ a.u.
(a) Monte Carlo simulation data calculated by Price and
Das Sarma \cite{rodney} (open circles), and the interpolation forms by Fano
and Ortolani \cite{fano} and Koskinen
{\em et al.} \cite{koskinen} (Eq. (\ref{koskinenexc})). The zero-field limit 
by AMGB is shown by the dotted line.
(b) Data of Price and Das Sarma \cite{rodney},
the Pad\'e approximant used by Ferconi and Vignale \cite{ferconi} 
(Eq. (\ref{eq:pade})) and the new fit (Eq. (\ref{eq:new})) to the data of
Price and Das Sarma $e^{\rm new}_{\rm xc}$.
}
\label{fig:exc}
\end{figure}

\pagebreak

\begin{figure}
\centerline{\psfig{figure=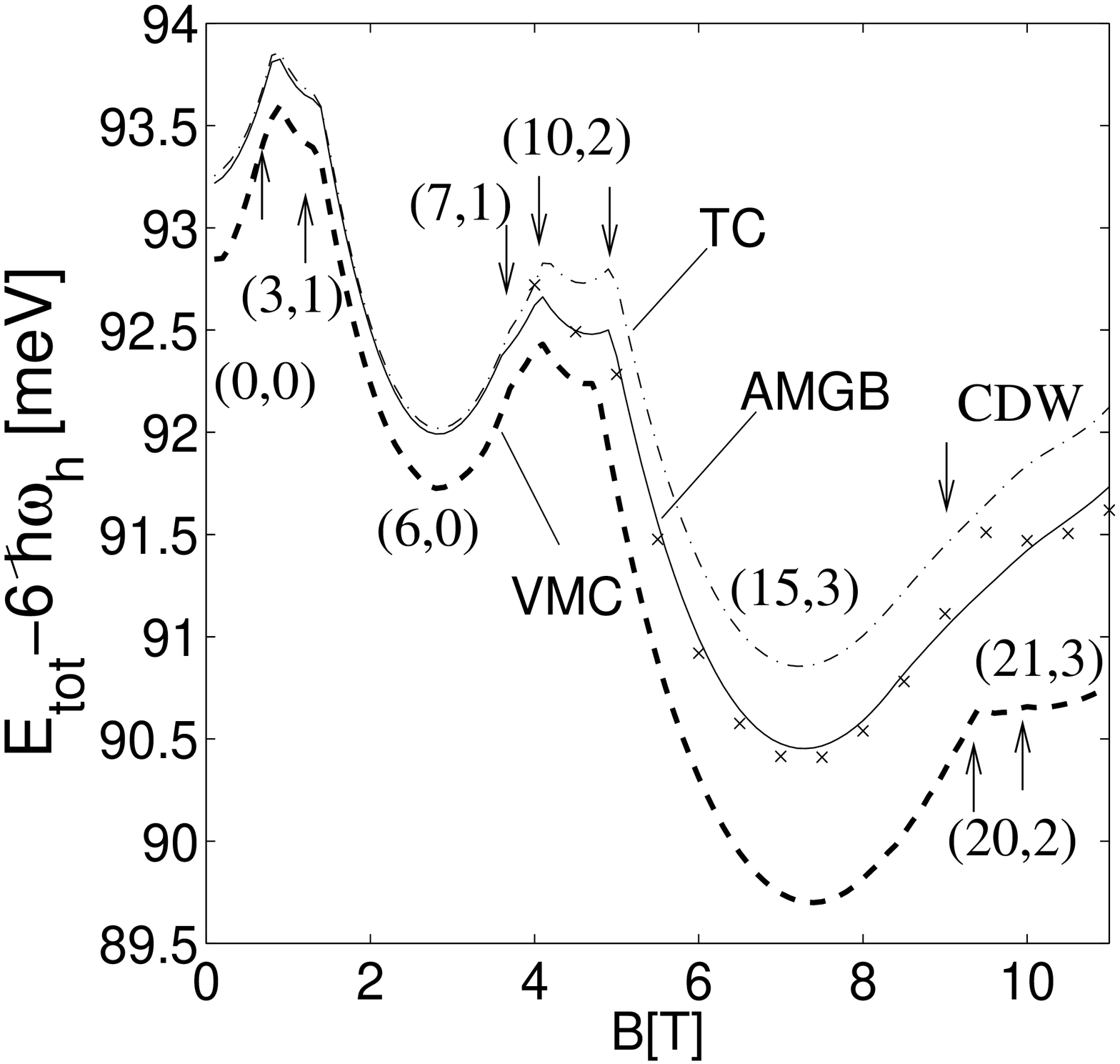,height=10cm}}
\centerline{\Huge\bf (a)}
\vspace{0.5cm}
\centerline{\psfig{figure=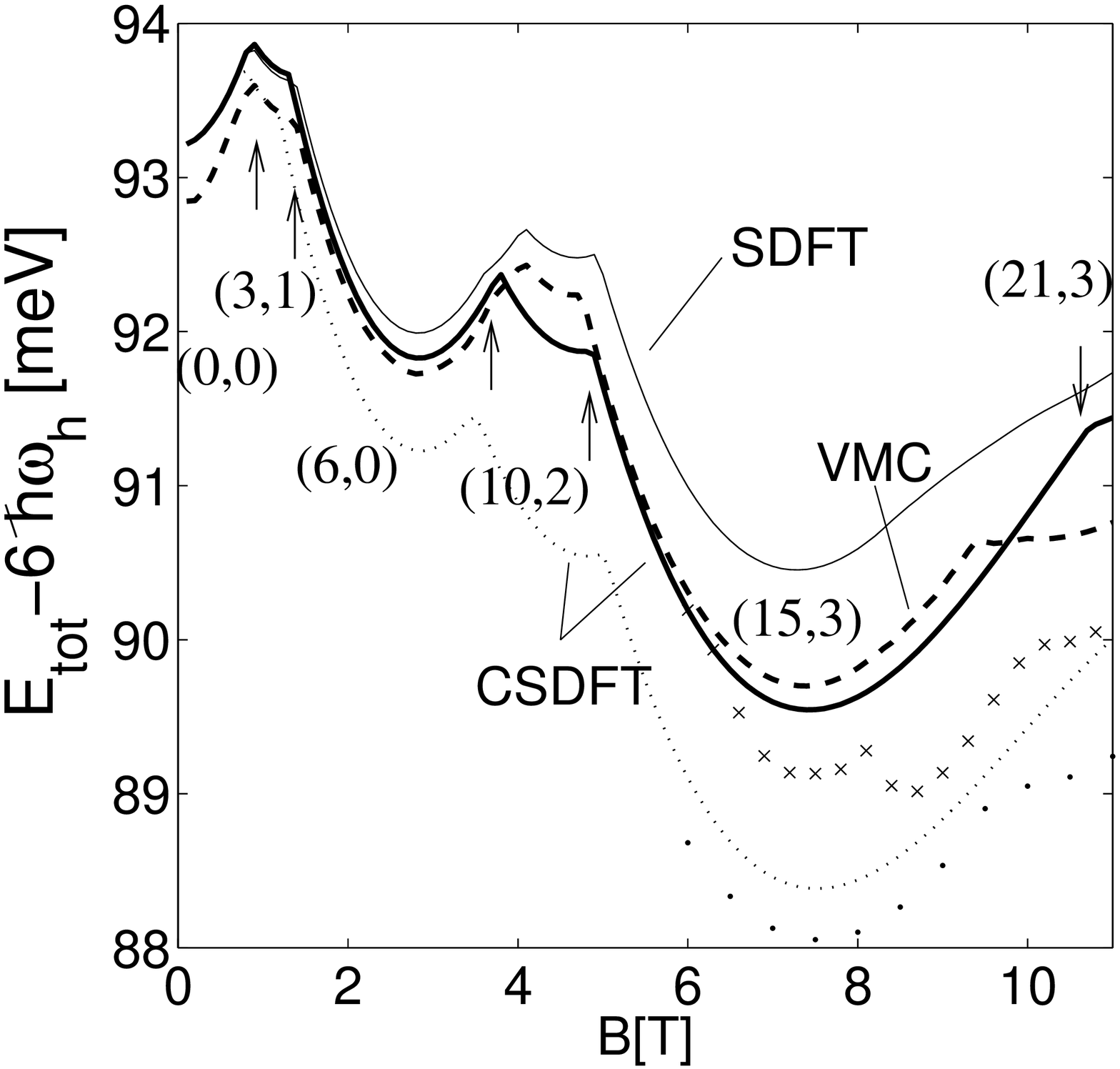,height=10cm}}
\centerline{\Huge\bf (b)}
\caption{
Ground state energy (minus $6\; \hbar \omega_h=6\; \hbar
\sqrt{\omega_0^2+(\omega_c/2)^2}$, where $\omega_c=eB/m^*$ is the
cyclotron
frequency) for the six-electron quantum dot in parabolic confinement
with the strength $\hbar\omega_0=5\textrm{ meV}$.
(a) SDFT results using the AMGB (solid line) and TC (dash-dotted line) 
functionals. The VMC results are marked with the bold dashed line.
The x-marks denote the CSDFT results obtained using the exchange-correlation
functional by Koskinen {\em et al.} (Eq. (\ref{koskinenexc})).
CDW denotes charge density wave state which breaks the rotational symmetry.
(b) CSDFT results using the exchange-correlation interpolations by
Ferconi and Vignale (Eq. (\ref{eq:pade})) (bold solid line) and 
the new functional  (Eq. (\ref{eq:new})) (dotted line). The
best SDFT results are repeated (solid line). The zero-field
AMGB exchange-correlation functional has been used in these calculations.
The x-marks and dots denote the CSDFT results for the $S_z$ = 2 state
obtained using the Ferconi and Vignale functional and the new functional,
respectively. The z-components of the total spin and and the total 
angular momentum are given in parenthesis as ($L_z$, $S_z$) and
arrows mark the transition points. It should be noted that the (7, 1) state
is missing in CSDFT.
}
\label{fig:magfield}
\end{figure}

\end{document}